\documentstyle[aps,preprint]{revtex}
\begin{document}

\draft

\title{Theory of arbitrarily polarized quantum Hall states}
\author{Sudhansu S. Mandal and V. Ravishankar}
\address{Department of Physics, Indian Institute of technology,
Kanpur -- 208 016, INDIA }

\maketitle

\begin{abstract}
We propose a global model which accounts for all the
observed quantum Hall states
in terms of an abelian doublet of Chern-Simons gauge fields,
with the strength of the Chern-Simons term given by a coupling
matrix. The model is employed within the composite fermion
picture.

\end{abstract}


%\newpage

The quantum Hall effects in two dimensional systems are observed
\cite{klit,tsui} at
high magnetic fields ($B \sim 10 $T).
In such a case, the spin degree
of freedom of the electron gets frozen in the direction of the magnetic
field (which is perpendicular to the plane of the system) and has no
dynamical role. Thus one assumes that the electrons are spinless for
these fully polarized quantum Hall (QH) states. The situation
changes, however, when $B$ becomes relatively small:
the Zeeman splitting is
now not that large (also partly due to small $g$ value, $g\sim 0.4$)
and, as Halperin \cite{halp} has observed, these systems are not fully
polarized. Indeed, experiments reveal that at relatively
small values of $B$, the QH states at filling factors $\nu = {4 \over
3}, \, {8 \over 5},\, {10 \over 7}$
\cite{clark,eisen1} and ${2 \over 3}$
\cite{eisen2,engel} are unpolarized
while the states at $\nu = {3 \over 5}$ \cite{engel} and ${7 \over 5}$
\cite{clark} are partially
polarized. Further, it is also known experimentally
that the states
which are at maximum polarization to start
with pass over to partially polarized
or unpolarized states as the Zeeman energy is lowered sufficiently ---
either by reducing the tilting angle of the magnetic field
\cite{clark,eisen1,eisen2,engel} or
by decreasing the electron density \cite{eisen2}.
In the vanishing Zeeman splitting (VZS)
limit, it has been found  from numerical computations \cite{tchak1}
that the states with
$\nu =2/(2n+1)$ are unpolarized and those of the Laughlin sequence
\cite{laugh} with $\nu =1/(2n+1)$ are fully polarized, in the
thermodynamic limit. Also the state at $\nu ={3 \over 5}$ has been
found to be partially polarized by an exact
diagonalization study \cite{tchak2},
in agreement with experiments.

Wu, Dev and Jain \cite{wu} have studied this problem and constructed
trial wave functions by employing the composite fermion model (CFM),
proposed originally by Jain \cite{jain}. These trial wave functions
are confirmed to be exact by numerical computation.
They report that, in the VZS limit, all even numerator QH states
are unpolarized and all those states with both the numerator and
denominator (of $\nu$) odd are partially/fully polarized. Further,
Belkhir and Jain \cite{belk} have proposed
that the CFM accommodates
the sequence $\nu =2n/(3n+2)$ all of which are spin unpolarized.
 From the wave functions that they construct, they also interpret that
these states possess a new feature where each composite fermion
carries two different types of vortices --- one of which seen by
all electrons while the other is visible only to an electron
of like spin.

We present here a generalized and consistent study of QH states
in an arbitrarily
polarized state within the frame work of Chern-Simons (CS) theory,
in the same spirit as the study of Lopez and Fradkin \cite{lopez} for
fully polarized states. To that end, we
introduce a doublet
of CS gauge fields
\begin{equation}
a_\mu = \left( \begin{array}{c}
    a_\mu^\uparrow \\
    a_\mu^\downarrow \end{array}
    \right) \; ,
\label{eq1}
\end{equation}
and let the strength of the CS term be matrix valued and have
the form
\begin{equation}
\Theta = \left( \begin{array}{cc}
    \theta_1 & \theta_2 \\
    \theta_2 & \theta_1 \end{array}
    \right)  \; .
\label{eq2}
\end{equation}

We adhere to the CFM, i.e., we insist that each electron have an
even number of vortices attached to it, although
the Aharonov-Bohm phase
picked up by one electron around the other
can be spin dependent.

The model described below bears a striking resemblence to a similar
model proposed recently by Lopez and Fradkin
(LF) \cite{lopez4} in order to
describe QH effect in double layered systems. In remarking so,
we emphasize that the models are {\it not} equivalent and the results
for the double layered systems do not go over
automatically to the spin systems
at hand here. Further elaboration will be taken up after we analyse
our model in detail.

Consider a two-dimensional system of non-relativistic spin-1/2 electrons
in the presence of magnetic field perpendicular to the plane. In their
study of spinless fermions, Lopez and Fradkin \cite{lopez} have shown
that such a system is equivalent to the one interacting with a CS
gauge field. Using this generic argument
we write a generalized Lagrangian
density as
\begin{equation}
{\cal L} = \psi_\uparrow^\dagger {\cal D}
     (a_\mu^\uparrow )\psi_\uparrow
   + \psi_\downarrow^\dagger {\cal D}(a_\mu^\downarrow )\psi_\downarrow
   +{1 \over 2} \tilde{a}_\mu\epsilon^{\mu\nu \lambda } \Theta
   \partial_\nu a_\lambda \; .
\label{eq3}
\end{equation}
Here $\psi$ is the fermionic field and
$\uparrow $ and $\downarrow $ represent
the spin up and down respectively.
${\cal D}(a_\mu^s) = iD_0^s +(1/2m^\ast)
D_k^{s\, 2} +\mu +(g/2)\mu_B (B+b^s)\sigma $
with $D_\mu^s =\partial_\mu -ie
(A_\mu +a_\mu^s)$ where $A_\mu $
is the external electromagnetic field and
$a_\mu^s$ is the CS field which interacts with the particles having spin
indices $s = \uparrow ,\, \downarrow $.
(We have chosen the units $\hbar = c=1$.)
The fixed density of particles
in the system is implemented by the
introduction of chemical potential $\mu$
which acts as a Lagrange multiplier.
$e$ and $m^\ast$ are the charge and effective mass of the
electron respectively.
Note that the Zeeman term comprises of
both the external magnetic field $B$
and the CS field $b^s$. $\mu_B$ is the
Bohr magneton, $\sigma = +1(-1)$ for spin-up(down) electrons and
finally,
$\tilde{a}_\mu $ is the transpose of the doublet field $a_\mu $.
The action given by Eq.~(\ref{eq3}) is invariant under the gauge
transformations $a_\mu^{ \uparrow , \downarrow } \rightarrow
a_\mu^{ \uparrow , \downarrow } + \partial_\mu \lambda ^{
\uparrow , \downarrow } (x)\, ,\, \psi_{ \uparrow , \downarrow }
(x) \rightarrow  \exp [ie \lambda^{ \uparrow  , \downarrow }
(x)] \psi_{ \uparrow  , \downarrow } (x)$. It should be
emphasized we are within the abelian CS theory.

The above Lagrangian density has several interesting features. Let
us diagonalize the matrix $ \Theta $, with eigen values $\theta_\pm
= \theta_1 \pm \theta _2 $, and denote $a_\mu$ in the eigen basis by
\begin{equation}
a_\mu =  \left(
\begin{array}{c}
a_\mu^+ \\
a_\mu^-
\end{array}
\right)  \; .
\end{equation}
By simple rescalings Eq.~(\ref{eq3}) may be written as
\begin{equation}
{\cal L} = \psi_\uparrow^\dagger {\cal D}
(a_\mu^+ +a_\mu^-)\psi_\uparrow
  + \psi_\downarrow^\dagger {\cal D} (a_\mu^+ -a_\mu^-)\psi_\downarrow
  +{\theta_+ \over 2 }\epsilon^{\mu\nu \lambda}
  a_\mu^+\partial_\nu a_\lambda^+
  +{\theta_- \over 2 }\epsilon^{\mu\nu \lambda}
  a_\mu^-\partial_\nu a_\lambda^-
  \; . \label{eq5}
\end{equation}
The relevant equations of motion now read
\begin{mathletters}
\begin{equation}
{\cal D}(a_\mu^+ +a_\mu^- \equiv a_\mu^\uparrow )
 \psi_\uparrow =0 \; \; ; \; \; \;
{\cal D}(a_\mu^+ -a_\mu^- \equiv a_\mu^\downarrow )
\psi_\downarrow =0 \; ;
\label{eq6a}
\end{equation}
\begin{equation}
\theta_+ b^+ = -e(\rho_\uparrow+\rho_\downarrow) \equiv -e\rho \;\;;
  \;\;\;
\theta_- b^- =-e(\rho_\uparrow-\rho_\downarrow)\equiv -e\Delta\rho \; ,
\label{eq6b}
\end{equation}
\end{mathletters}
where $\rho_\uparrow (\rho_\downarrow) $ is the density
of spin up (down) particles.
It is clear that the above equations naturally incorporate the idea
that each electron has two kinds of vortices associated with it.

Case--I: We first study the case $\theta_1 = \theta_2 $.
Here $\theta_+ = 2 \theta $ and $ \theta_- \equiv 0$. Thus the
CS gauge field $a_\mu^- $ decouples dynamically and merely plays
the role of a Lagrange multiplier: ${\partial {\cal L} \over
\partial a_0^- }= \Delta\rho \equiv 0$. Thus the unpolarized
case is accomplished by the choice $\theta_1 = \theta_2 $.
Rescaling $\theta $ by ${\theta \over 2}$, we parametrize
$\theta = {e^2 \over 2\pi}({1 \over 2s})$ ($s$ is an integer) in
order to impose composite fermion picture on the model. Note
that the ensuing CFM is different from the one envisaged by
Belkhir and Jain \cite{belk} who distinguish between the
relative phase between like spin particles and unlike spin
particles for singlet states.
In this model, there is no such phase distinction.

The determination of the filling fractions is now straight
forward. Since there is only one CS gauge field $a_\mu^+$, a
standard mean field (MF) ansatz leads to an average CS magnetic
field $\langle b^+ \rangle = -{e\rho \over \theta }$ which is
seen by {\it all} the electrons. Demanding that the
effective Landau levels (LL) formed by the effective magnetic
field $\bar{B}^+ = B +\langle b^+ \rangle $ accommodate all the
particles at an integer filling factor $2p$, ($p$ for up
spin and $p$ for down), the true filling fraction $\nu$ is
obtained as
\begin{equation}
\nu = {2p \over 4sp +1} \; .
\label{eq7}
\end{equation}
The energy corresponding to each level is obtained as
$\varepsilon_{n\sigma} =(n+1/2)\bar{\omega}_c -{g \over 2} \mu_B
\bar{B}^+ \sigma$ ($n=0,1,\ldots $),
where the effective cyclotron frequency
$\bar{\omega}_c ={e \over m^\ast}\bar{B}^+$.
All the states obeying Eq.~(\ref{eq7}) are spin unpolarized.
Recall that $p$ can be a negative integer (meaning $\bar{B}^+ $
is antiparallel to $B$). Note that the sequence (\ref{eq7}) is
exactly the same that was obtained by Wu et al. \cite{wu}, and
does indeed accommodate all the known experimentally observed
states and also maintain consistency with the numerical result
that all even numerator states are unpolarized. In the limit $p
\rightarrow  \infty$, $\nu \rightarrow  1/(2s) \Rightarrow $
that all even denominator states are also unpolarized. Further
by particle-hole symmetry, the states $2-\nu $, and
the states $2+\nu $ which are obtained by the addition of LL
\cite{wu}, are also unpolarized.
It is indeed true that the even-numerator levels such as $\nu
={4 \over 3}$, ${8 \over 5}$ and ${10 \over 7}$ \cite{clark}
and even-denominator state like $\nu
={5 \over 2}$ \cite{expt1} have been experimentally observed to
be unpolarized.

Case--II: We now consider the case $\theta_1 \neq \theta_2 $ in
order to obtain partially polarized states. In the diagonal
basis $\theta_\pm = \theta_1 \pm \theta_2 \neq 0$. The study of the
Lagrangian
now offers several novel features. The phase picked up by a spin
around a like-spin is different from what it would pick up around
an unlike-spin. Moreover, the MF ansatz now involves smearing two
fields $b^\pm $ which are unequal. We find that
\begin{equation}
\bar{B}^\uparrow = B+b^+ +b^- \; \; ; \;\;\;
\bar{B}^\downarrow = B+b^+
-b^-  \label{eq8}
\end{equation}
are respectively the fields seen by spin up and spin down
particles. Consequently there are two energy scales
corresponding to the two gaps at the MF level. We again
implement the composite fermion requirement.
This leads to the condition
\begin{equation}
-e\left( {1 \over \theta_+} +{1 \over  \theta_-}
\right)=-{2\pi \over e}(2s) \; .
\label{eq9}
\end{equation}
However allowing the unlike spins to pick up fermionic as well
as bosonic phases we get
\begin{equation}
-e\left( {1 \over \theta_+} -{1 \over \theta_-}
\right)=-{2\pi \over e} k \; ,
\label{eq10}
\end{equation}
where $k$ is an arbitrary integer.

Let us parametrize $\theta_\pm = (e^2 / 2\pi )(1 / s_\pm
)$. Demanding that exactly $p_\uparrow (p_\downarrow )$
numbers of effective LL be filled by spin up (down) electrons,
we obtain
\begin{mathletters}
\label{eq11}
\begin{eqnarray}
{\rho_\uparrow \over p_\uparrow} &=& {\rho \over \nu}- \left[
\rho s_+ + (\Delta\rho)s_- \right] \; ,  \\
{\rho_\downarrow \over p_\downarrow } &=& {\rho \over \nu}- \left[
\rho s_+ - (\Delta\rho)s_- \right] \; .
\end{eqnarray}
\end{mathletters}
Note that if $\Delta\rho =0$, $s_-$ is irrelevant. In other
words, the requirement of unpolarized spin states causes a collapse
to Case-I. Solving for $\Delta\rho$ and $\nu $ we obtain
\begin{mathletters}
\label{eq12}
\begin{eqnarray}
{\Delta\rho \over \rho} &=& {p_ \uparrow -p_ \downarrow  \over p_
\uparrow +p_ \downarrow +4s_- p_ \uparrow p_ \downarrow }\; ,\\
\nu &=& { p_ \uparrow +p _ \downarrow
+4s_-p_ \uparrow p_ \downarrow  \over
1 + (s_+ +s_-)(p_ \uparrow +p_ \downarrow )+ 4s_+ s_- p_ \uparrow
p_ \downarrow } \; .
\end{eqnarray}
\end{mathletters}
Observe that $\Delta \rho \neq 0 $ if and only if
$p_ \uparrow \neq p_ \downarrow $.
Note that for  $\Delta \rho =0 $,
$s_- $ (being an irrelevant parameter)
needs to be dropped.
It is not too tedious to verify from a one-loop computation that
the Hall conductivity gets quantized at precisely those values
of $\nu $ (given by Eqs.~(\ref{eq7}, and \ref{eq12})), subject to
the weak restriction $\lim_{{\bf q}^2 \rightarrow  0} V({\bf
q}^2 ){\bf q}^2 =0$, where $V({\bf q}^2)$ is the electron
interaction potential.

We study  the extreme case $s_- =0$, i.e., $\langle b^-
\rangle =0 $ first. The CS field $a_\mu^- $ is decoupled at the tree
level (in contrast to the unpolarized states where $a_\mu^- $ is
completely non-dynamical). Consider then the sequence $p_ \uparrow =
p_ \downarrow +1 =p$. Then
\begin{equation}
{\Delta \rho \over \rho }={1 \over 2p-1}\;\; ; \;\;\; \nu = {2p -1
\over s_+ (2p-1)+1} \; .
\label{eq13}
\end{equation}
We have $s_+ =2s $ by virtue of composite fermion
requirement. Thus
\begin{equation}
\nu = {2p-1 \over 2s(2p-1)+1} \; .
\label{eq14}
\end{equation}
These states are indeed partially polarized becoming fully
polarized when $p=1$. Then $\nu_{p=1}= 1/(2s+1)$ which is simply
the Laughlin sequence \cite{laugh} known to be completely
polarized \cite{tchak1}. The case $s=1$ yields the sequence
obtained by Wu et al. \cite{wu}. Particle-hole symmetry and the
addition of LL imply again that $2-\nu$ and $2+\nu$ are also
spin-polarized. It turns out that the sequences given by
Eqs.~(\ref{eq7} and \ref{eq14})
exhaust all known integer and fractional
states --- with full, partial or no polarization.

The model can accommodate many more states corresponding to
$\bar{B}^\uparrow \neq \bar{B}^\downarrow $. As an interesting
exercise, let us
construct the sequence of states that would
follow from Belkhir and Jain proposal \cite{belk}. This
corresponds to the choice $k=2s-1$ in (\ref{eq10}),
whence, $s_+ =(4s-1)/2 $
and $s_- =1/2$. We obtain
\begin{mathletters}
\label{eq15}
\begin{eqnarray}
{\Delta\rho \over \rho} &=& {p_ \uparrow - p_ \downarrow \over p_
\uparrow + p_ \downarrow +2p_\uparrow p_\downarrow } \; ; \\
\nu &=& {p_ \uparrow + p_ \downarrow +2p_ \uparrow p_ \downarrow
\over 1+ 2s(p_ \uparrow +p_ \downarrow ) +(4s-1)p_ \uparrow p_
\downarrow } \; ,
\end{eqnarray}
\end{mathletters}
which clearly is a new sequence.

In general, the family of sequences is given in terms of
four independent
parameters $(p_ \uparrow \, ,\, p_ \downarrow \, ,$ $ s_+ ,\, s_- )$
subject to the composite fermion constraint ---- $ s_+ +s_- = 2s$
(an even integer). It raises the interesting question of the
uniqueness of the $\nu $ values obtained. We note that the sequences
are indeed different, and any accidental degeneracy (i.e., same
$\nu $ value from different sequences) does not make the model
ambiguous. For, it would correspond to different gap energies,
which can be determined, say, by the activation of diagonal
resistivity.
In short, the QH states are labeled by both $\nu $ as well as
$\bar{\omega}_c^{ \uparrow , \downarrow }$. It is interesting
that all states that are known so far correspond to $\bar{\omega}_c^
\uparrow = \bar{\omega}_c^\downarrow $.

It remains to contrast our results with those of LF \cite{lopez4}
who studied double layered systems. As we observed, there is a close
resemblance between the two models: The Lagrangians are formally
the same, with a fermion doublet and a matrix valued CS strength.
The crucial difference, however, is in the (physical) choice of
$\Theta$. While $\Theta_{{\rm LF}}^{-1}$ is well defined in
Ref.~\cite{lopez4}, it is $\Theta$ that is so in our case. This leads
to certain fundamental differences in results
and interpretation which we
list below:\\
(i) Consider the spin unpolarized state first. The corresponding
sequence of states obtained by LF \cite{lopez4} for equal
population in the two layers is identical to Eq.~(\ref{eq7}) here.
A closer look however shows that it precisely for these states,
(characterized by the filling fractions $\nu_1 =\nu_2$ in  the two
layers and the number of particles $N_1=N_2$ in two layers in
Ref.~\cite{lopez4}) that the CS strength $\Theta_{{\rm LF}}$
becomes ill-defined. The ensuing dynamics is also ill-defined.
Indeed as LF \cite{lopez4} point out in their paper ``the spin
singlet state (3,3,2), which has filling fraction $\nu =2/5$,
cannot be described within the Abelian Chern-Simons approach".
Observe that, in contrast, our $\Theta$ is well-defined (albeit
$\Theta^{-1}$ is not, but that is irrelevant), and we have shown
that the choice $\theta_1 =\theta_2$, naturally leads to
unpolarized states. The dynamics is also well defined, allowing
us to perform the standard one-loop computation to verify
quantization of Hall conductivity $\sigma_H$ at these filling
fractions and also derive the many-body wave function for
singlet states thereof \cite{wfn}.
(ii) Consider the states $\nu = 1/m$. In our approach, odd $m$
always corresponds to fully polarized states while $m$ even
can be a spin unpolarized state. Contrarily in
Ref.~\cite{lopez4}, one can have (analogues of) unpolarized
states for all m (even or odd). Clearly, our choice is
closer to the experiment \cite{engel} and numerical calculations
\cite{tchak1} in a single layer.
(iii) Consider the state $\nu =1/2$.
For their physical states (i.e., in agreement with
numerical computation), LF \cite{lopez4} assign
$\nu_1 =\nu_2 =1/4$ yielding gaps $\bar{\omega}_c^{1,2}
=\omega_c /4 $ (where 1 and 2 refer to two different layers).
The present model, in contrast, yields $\bar{\omega}_c^{\uparrow ,
\downarrow} =0$, which is again closer to the experiments
\cite{will,kang,gold} which have verified that $\bar{\omega}_c =0$
for $\nu =1/2$ in a single layer.
(iv) It is not that the two models are entirely dissimilar in their
prediction. Indeed, the double layered systems are exact analogues
of the spin systems with dissimilar gaps for spin up and down
states. Note that spin unpolarized states (which are seen
experimentally \cite{clark,eisen1,eisen2,engel}) can never arise
if $\bar{\omega}_c^\uparrow \neq \bar{\omega}_c^\downarrow $.
Experimentally observed partially polarized or fully polarized
states also correspond to $\bar{\omega}_c^\uparrow
 = \bar{\omega}_c^\downarrow $. (v) Finally, we remark that LF
\cite{lopez4} have also studied spin unpolarized states by employing
a non-abelian CS interaction, with additional new features such
as semion statistics obeyed both by the charged spinless holons
and neutral spin-1/2 spinons --- which indicates a departure
from CFM to which we completely adhere. A comparision of this
SU(2) model with the abelian formalism developed here merits
further study.

To conclude, we have developed a global model for all the
observed QH states in terms of a doublet of CS fields with the
coupling matrix of form (\ref{eq2}). Our findings are consistent
with the exact results obtained numerically as well as by
experimental observations.
Finally, there remains the determination of the wave functions
for these states from the Lagrangian (\ref{eq5}) following the
beautiful method developed by Lopez and Fradkin \cite{frad}.
We report
here that for the state $\nu = {2 \over 5 }$, the Halperin
wave function \cite{halp} is remarkably recovered, as a
striking vindication of the model. The details will be reported
else where \cite{wfn}.  \\

We thank the referees for their insightful queries and suggestions.

\end{document}